\documentclass[printer]{aa}
\usepackage{txfonts}
\usepackage{graphicx}
\usepackage{natbib}
\bibpunct{(}{)}{;}{a}{}{,}
\begin{document}

\title{A photo-ionized canopy for the shock-excited Criss-Cross
Nebula \thanks{Based on observations made with the Nordic Optical
Telescope, operated on the island of La Palma jointly by Denmark,
Finland, Iceland, Norway, and Sweden, in the Spanish Observatorio
del Roque de los Muchachos of the Instituto de Astrofisica de
Canarias. }}

\author{S. Temporin\inst{1,2}, R. Weinberger\inst{2} \and B. Stecklum\inst{3}}

\institute{CEA/DSM/DAPNIA, Service d'Astrophysique, Saclay, 91191 Gif-sur-Yvette Cedex, France
\email{sonia.temporin@cea.fr} \and
Institut f\"ur Astro- und Teilchenphysik, Technikerstra\ss e 25,
6020 Innsbruck, Austria\\ \email{Ronald.Weinberger@uibk.ac.at} \and
Th\"uringer Landessternwarte Tautenburg,
          Sternwarte 5, 07778 Tautenburg, Germany\\
             \email{stecklum@tls-tautenburg.de}}

\date{Received 20 December 2006; accepted 07 February 2007}

%\abstract % max 300 words, no references, no abbreviations or acronyms
%{}{}{}{}{}
\abstract{}{We study a new broad well-defined arc of optical
nebulosity close to the cloud-shock interacting Criss-Cross
Nebula, derive the basic physical properties of the former and
revise those of the latter, and compare both objects to
simulations of cloud-shock interactions from the literature.}
{Deep optical, partly wide-field, images were used to reveal the
intricate morphology and overall extent of the nebulosities.
Optical spectroscopy enabled us to uncover their nature. }{The two
nebulosities obviously are physically linked, but are of different
type; the Criss-Cross Nebula is, as was shown also in an earlier
paper, excited via a slow shock from the expanding Orion-Eridanus
Bubble, but the broad arc is definitely photo-ionized. The source
for ionizing photons appears to be hot gas in this bubble. Some
results of simulations of interactions of SNRs with interstellar
clouds available from the literature bear a striking resemblance
to our nebulae, which appear to represent an example - unrivalled in
closeness and clarity  - for an early to medium stage in the
destruction of an isolated cloud over-run by a highly evolved SNR.
Thereby the Criss-Cross Nebula is, when seen from the SNR, the
rear disrupted part of the original small cloud, whereas the arc
probably is its yet rather intact front part. }{}

 % max. 6 key words

\authorrunning{S. Temporin et al.}

\maketitle
%\titlerunning{}

\section{Introduction}
The interstellar medium (ISM) is a highly turbulent compound of
various interacting gas phases and of dust. Supernova explosions
probably dominate the global kinetic energy input into the ISM
\citep{lk04}. Their remnants (SNRs) are ubiquitous in the Galaxy
and the majority of these estimated 20\,000-30\,000 objects are
very old and visible in \ion{H}{i} only, but merely about 230
Galactic SNRs, i.e. younger examples of their kind, are currently
known \citep{kk06}. The old SNRs fill large volumes of space; many
condensations (clumps, clouds) in the ISM will be created
and/or shaped by them and by the stellar winds of their
predecessors, and a number of clouds will eventually be destroyed
by shock-cloud interactions. Despite the frequent occurrence of
the latter processes and their obvious significance for
understanding the dynamics, the mass distribution within the ISM
and the mixing and evolution of the ISM, only about a dozen
interactions between (isolated) clouds and shocks are known. On
the other hand, there are numerous theoretical - and recently,
also one major experimental - studies on this topic \citep{sn92, km94, op05,
bk06, nmkf06, kb03}. With respect to observed SNR shock-cloud
interactions, the Cygnus Loop SNR turned out to be particularly
prolific: because of its large angular size
(3$\fdg5\times2\fdg8$), low foreground extinction (a few tenths of
mag in $A_{\rm V}$) and various shock conditions, it serves as a
favorable location for investigating shock-cloud interactions of
middle-aged remnants \citep{pf05}.

Evolved, very close SNRs and/or wind-blown (super)-bubbles, located
at moderate to high Galactic latitudes would obviously be optimum
for studying how the majority of SNRs interacts with (isolated)
ISM clouds or clumps. Due to the rather small expansion velocities
of objects of this type (less than $\sim$100 km s$^{-1}$), one
would neither expect X-ray emission nor highly-ionized atomic
species from such shocked clouds. Instead, the latter would show
up as small optical nebulae of complex morphology and display an
emission spectrum  typical for shock-excitation (e.g. prominent
[\ion{S}{ii}] lines).

The Orion-Eridanus (Super)Bubble (OEB) could be a promising target
for a search for interactions with ISM clouds. Its size (about
45$\degr\times35\degr$) is comparable to its distance ($\sim$0.25
kpc) and its centre is at ($\ell,b$) $\approx$
(195$\degr$,$-35\degr$). The OEB is filled with hot gas ($T
\sim10^6$ K) and is bounded by a slowly expanding ($v \approx 40$
km s$^{-1}$) \ion{H}{i} shell, the near side of which is only
$\sim$0.15 kpc away from us \citep{hh99, gb95, bh95}. The total
mass of the \ion{H}{i} shell is estimated to be about 2.3 10$^5$
$M_{\odot}$, and its kinetic energy about 3.7 10$^{51}$ erg
\citep{bh95}. \citet{hh99} call the OEB the `Rosetta Stone` of
superbubbles, being in a middle evolutionary stage, having
originated long ago and still being energized by massive stellar
winds and supernovae, and exhibiting a remarkable range of
astrophysical processes.

One decade ago, in a preliminary study we presented a new faint
small (6$\arcmin \times 3\arcmin$) optical nebula of curious
filamentary morphology that was found to be projected against the
centre of the OEB \ion{H}{i} shell \citep{zw97}. Because of its
unique shape we dubbed it `Criss-Cross Nebula` (CCN). The object
had no known counterpart at any other wavelengths.

 A medium
resolution and signal-to-noise optical spectrum (4200--6900 \AA )
showed a few emission lines from low ionization species only
([\ion{O}{iii}] was not detected): the line ratios
H$\alpha$/[\ion{S}{ii}] =  0.9 and H$\alpha$/[\ion{N}{ii}] = 1.2
prompted these authors to classify it as probably being
shock-excited. They concluded that the CCN is a small, perhaps
isolated, cloud over-run by the slow moving shock wave responsible
for the OEB and argued for a distance of 150 pc for the CCN. The
CCN remained practically unnoticed in the subsequent years. Very
recently, however, \citet{ms06} detected the CCN in both near and
far ultraviolet passbands and modelled the UV emission as
two-photon emission from a moderate velocity shock.

In 1999 we obtained a (to date unpublished) image of the CCN that
reveals a complex network of filaments. Hitherto, there were only
reproductions from the  Palomar Observatory Sky Survey (POSS)
available. However, the main reason why we are again dealing with
the CCN is the advent of wide-field, high-resolution CCD imaging
within the last decade, which  enable us to look for previously
unknown structures around the Criss-Cross Nebula and thus
to perform a more detailed comparison to results of numerical
simulations of cloud-shock interactions from the literature.

\section{Observations}

\subsection{Imaging}

An $R$-band 1200 s exposure covering a field of view of
$\sim$6\arcmin$\times$6\arcmin\ was obtained in 1999 November at
the Nordic Optical Telescope (NOT) with the faint object
spectrograph and camera ALFOSC. The image has a pixel-scale of
0\farcs188\ pixel$^{-1}$ and the seeing during the observations
was $\sim$ 0\farcs8. A \citet{la92} standard field for photometric
calibration was observed with the same filter three times during
the night. The standard reduction steps (bias subtraction,
flat-fielding, cosmic-ray cleaning), as well as the determination
of the magnitude zero-point and atmospheric extinction coefficient
through aperture photometry of the photometric standard stars were
performed with the available IRAF\footnote{IRAF is distributed by
the National Optical Astronomy Obser\-vatories, which are operated
by the Association of Universities for Research in Astronomy,
Inc., under cooperative agreement with the National Science
Foundation.} packages.

Broad band I (2$\times$180 sec) and narrow band H${\rm \alpha}$
(2$\times$1200 sec) and [\ion{S}{ii}] $\lambda$ 6716+6731 \AA\
(2$\times$1200 sec) exposures were acquired in 2005, October 8/9,
at the Th\"uringer Landessternwarte Tautenburg (Germany), with the
2\,m telescope in the Schmidt mode, for which the correction plate
limits the aperture to 1.34 m. The bandpass width (FWHM) of the
H${\rm \alpha}$ and [\ion{S}{ii}] filters is 97 \AA. The pixel
size of the 2k$\times$2k SITe CCD chip, 24$\times$24 $\mu$m,
and  the plate scale of 51\farcs4 mm$^{-1}$ give a field of view 
of 42\arcmin$\times$42\arcmin\,with an image scale of 1\farcs23
pixel$^{-1}$.

\subsection{Spectroscopy}

A long slit spectrum in the wavelength range $\lambda$ 6250 -
8190, with dispersion 1.93 \AA\ pixel$^{-1}$ and resolution $\sim$
9.5 \AA\ at $\sim$ 6600 \AA, was obtained with a 1 hr exposure in
2005, November 9, at the 1.82\,m telescope of the Padova
Astronomical Observatory at Asiago (Italy), equipped with the
Asiago Faint Object Camera and Spectrograph (AFOSC), that,
combined with a TEK 1024 thinned CCD chip, provides a spatial
scale of 0\farcs47 pixel$^{-1}$.

The spectrophotometric standard star G191B2B was observed
immediately after the target spectrum for flux calibration
purposes. The standard reduction steps (bias subtraction,
flat-fielding, wavelength linearization, sky-background
subtraction, atmospheric extinction correction and flux
calibration) were carried out with IRAF. Cosmic-ray events were
rejected by means of the package {\sc L.A.COSMIC} \citep{vd01}.

\begin{figure}
\centering
\includegraphics[width=8.7cm]{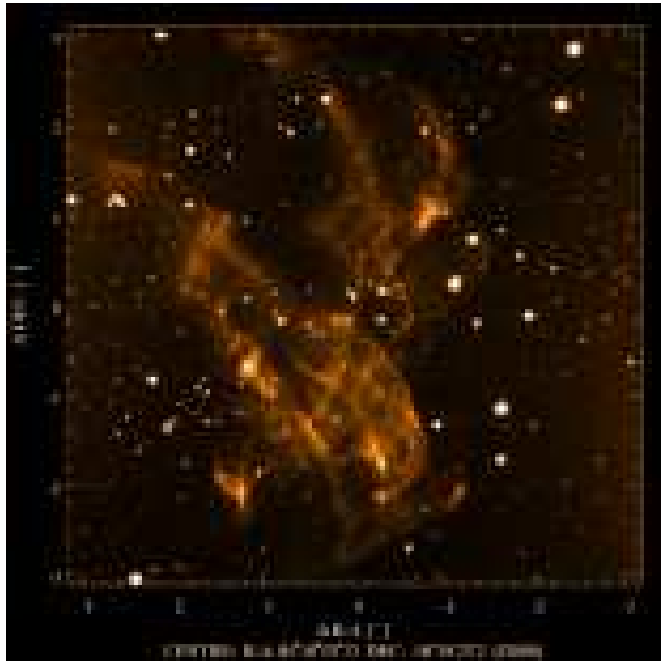}
\caption{Broad-band $R$ exposure of the Criss-Cross Nebula,
obtained with the faint object spectrograph and camera ALFOSC on
the 2.56\,m Nordic Optical Telescope (NOT).}
 \label{CC_fig_not}
\end{figure}

\section{Images of the CCN and its neigborhood}
The image taken at the NOT (Fig.~\ref{CC_fig_not}) 
shows the structure of the CCN in
considerable detail. Compared to the extent as visible
on the POSS, only a very small portion, namely its northernmost
tip (0\farcm3) is not covered by this image. The CCN appears to
consist of two main parts, a northern and a southern one,
connected by at least one long ($\sim4\farcm5$), slightly bent,
thin ($0\farcm1-0\farcm3$) filament, stretching N-S. The northern
part of the CCN appears to be made up of a part of the just
mentioned filament plus an additional western, more bent, filament
which seems to consist of at least two subfilaments and contains
one brighter region. This part of the CCN is about half as broad
in E-W direction compared to the southern part. The latter
stretches from NE to SW, with an extent of $\sim5\farcm5$ and
shows, particularly in its southern portion, a wealth of
criss-cross filaments and brightness enhancements, which partly
originate in the crossing of two filaments. The brightest spot
however, located at R.A. = 4$^{\rm {h}}$10$^{\rm {m}}$12.4$^{\rm
s}$, Dec. = $-$04\degr59\arcmin03\arcsec\ (2000), i.e.
$\sim1\farcm3$ SE of the centre of the image, does not seem to be
an outcome of such a projection. The apparent centre of the CCN,
1\farcm1 away from this bright spot towards the NW direction,
looks like a tiny horizontally extending rectangle with dimensions
0\farcm2$\times0$\farcm1 and has a 2000.0 position R.A. = 4$^{\rm
{h}}$10$^{\rm {m}}$08.2$^{\rm {s}}$, Dec. =
$-$04\degr58\arcmin33\arcsec\ ($\ell$ = 197\fdg0, $b$ =
$-37$\fdg8).

\begin{figure*}
\centering
\includegraphics[width=17cm]{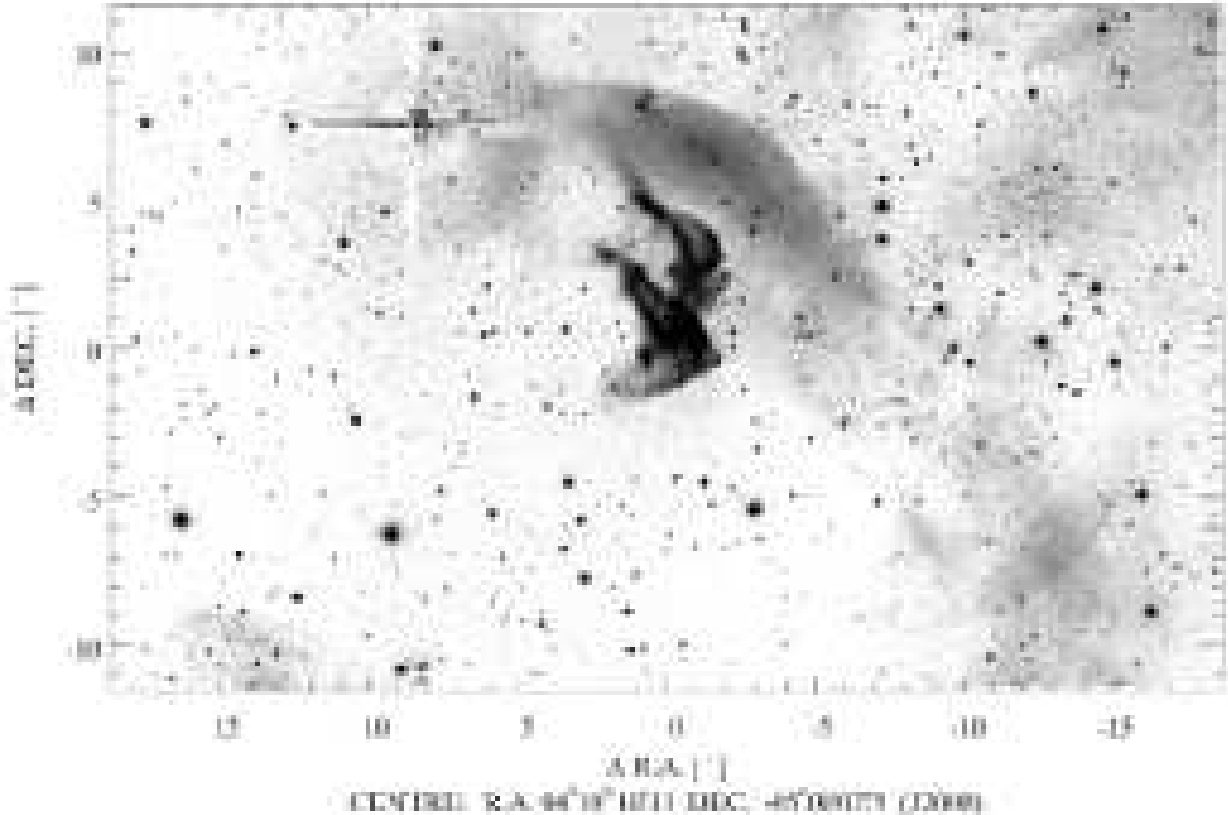}
\caption{Narrow-band wide-field H$\alpha$ exposure centered on the
CCN, obtained with the 2\,m telescope of the Th\"uringer
Landessternwarte Tautenburg in the Schmidt mode.}
\label{CrissCross_wide}
\end{figure*}

In Fig.~\ref{CrissCross_wide} we present a wide-field image taken
in the Schmidt mode of the 2\,m telescope of the Th\"uringer
Landessternwarte Tautenburg, centered on the CCN. The faint large
(16\arcmin \, in diameter) circle in the western part of the field
as well as the smaller circles or disks plus the dark horizontal
streak in the NE are residuals from removing bright stars and
hence are artificial.

 To the NW of the CCN, about 80\arcsec\ apart,
an extended broad arc with clear outer borders becomes visible. 
It appears to be a $\sim90\degr$
long portion of a circular structure with a diameter of
$\sim$20\arcmin, but there are no traces of this arc visible in
the W, SE or SW of the CCN. The centre of this circle would be
located very close ($\sim$1\arcmin), just outside, to the
south-easternmost part of the CCN. The width of this arc-like
structure is about $\sim$2\farcm7. There is an extension of the
northernmost filament of the CCN which seems to be
connected to the eastern part of the arc, at least in projection.

The CCN interestingly displays very sharp borders in this deep
wide-field image, particularly to the south and west, but shows -
with exception of its northernmost filament - no particular
additional emission outside the CCN compared to its extent on the
POSS or on the NOT image. However, in the N and, especially, NW
and W part of the field presented in Fig.~\ref{CrissCross_wide},
very faint nebulae are present: any connection to the CCN or the
arc is not obvious, but cannot be ruled out. Using the whole
field of view of 42\arcmin$\times$42\arcmin and very high contrast
does not change this perception. A thorough inspection of a deep
H{\rm$\alpha$} image of the OEB (Figs. 2 and 3 in \citet{bd01})
shows that the CCN is close to, but just outside, a faint  H{\rm
$\alpha$} protrusion of their 'Arc A`, the major arc of the
complex filamentary nebulosity first discovered by \citet{mea65,mea67}. 
In our Fig.~\ref{CrissCross_wide} the division of the CCN into two main
components, a northern and a southern one, is quite evident and
the large-scale direction of the northern parts of both components
are approximately parallel to the arc.

\section{Spectroscopy of the CCN and the arc}
\begin{figure}
\centering
\includegraphics[width=8.7cm]{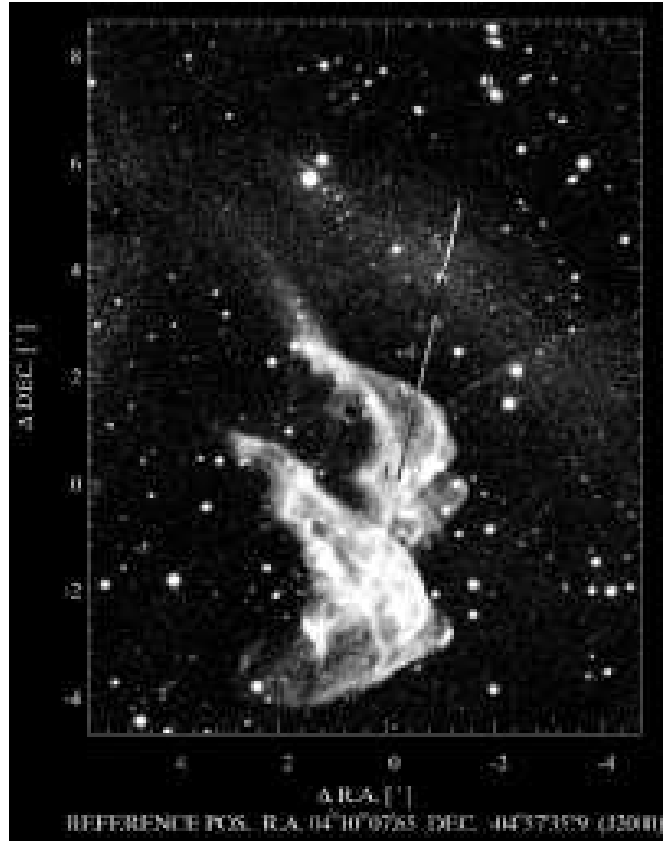}
\caption{Narrow-band [\ion{S}{ii}] image, on which the position of
the 7\farcm9-long spectroscopic slit, centered at R.A. = 4$^{\rm
{h}}$10$^{\rm {m}}$04$^{\rm {s}}$, Dec. =
$-$04\degr50\arcmin01\arcsec\ (2000), has been marked. Labels
indicate the regions whose emission line ratios are listed in
Table~\ref{emlines}. Regions 1-3 are within the CCN, region 4 and,
particularly, 5 belong to the broad arc.} \label{slit_pos}
\end{figure}

The 7\farcm9 long and 2\farcs1 wide slit was positioned in such a
way to cross some filaments in the CCN and the extended emission
feature (the arc) north  of it, as well as to include some
sky-background free of emission, as shown in Fig.~\ref{slit_pos}.
We used the [\ion{S}{ii}] image, because it shows much better
which parts of the inner structure of the CCN are covered by the
slit, but the broad arc is only faintly visible.

The emission lines H${\rm \alpha}$, [\ion{N}{ii}] $\lambda$ 6548
and 6583, [\ion{S}{ii}] $\lambda$ 6716 and 6731 were detected in
the regions of the CCN and  - with exception of [\ion{S}{ii}]
$\lambda$ 6731 -  also in the broad arc, although in the latter
region the forbidden emission lines are considerably weaker. We
obtained emission-line ratios through Gaussian fitting of the
aforementioned lines in 5 regions along the slit. The relevant
values with the associated relative errors are given in
Table~\ref{emlines} together with the extent of each region in
arcsec along the slit. The region labelled '5' and largely also
'4' identify the part of the extended region north of the CCN
(i.e. the broad arc).

In Table 1 a striking difference between the
[\ion{N}{ii}]/H$\alpha$ and the [\ion{S}{ii}]/H$\alpha$ ratio of
the CCN and those of the broad arc is obvious and will be discussed
below. This discrepancy prompted us to check whether  it is of
general nature or valid for a restricted area only. For this
purpose we constructed an H$\alpha$/[\ion{S}{ii}] image (not
presented here). This image unambiguously shows that the
discrepancy is present everywhere in the two nebulae. Only in the
northernmost portion of the northern filament, which appears to be
a link to the arc, there is a distinct gradient visible, smoothly
varying from the high value of the CCN to the low one of the arc.

A high-dispersion long-slit spectrum would be quite useful 
to irrevocably identify the nature of the CCN, and is planned for
follow-up observations of this object.
However, the radial velocity information derived from our low-dispersion 
spectrum gives us already some important clues to the nature of the CCN and
the northern arc. Radial velocities were obtained from the positional 
measurements of the emission-lines after splitting the spectrum in two parts, 
one including regions 1-3 and the other including regions 4-5. 
The wavelength calibration was based on the night-sky lines from 6250 to 6875 \AA.
The radial velocities of the two groups of regions are 
the same within the errors (estimated to be in the range 5-10 km s$^{-1}$ by
taking into account a slight dependence on the choice of the wavelength region for the
calibration) and are close to zero.  
Average values are v$_{\rm {LSR}}$ = 4.4 km s$^{-1}$ and
$-$ 1.4 km s$^{-1}$ for the CCN and the arc, respectively.
This is consistent with the two structures being inter-related local
phenomena.

\begin{table*}
\caption[]{Emission line fluxes} \label{emlines}
$$
\begin{tabular}{llllll}
\hline \hline \noalign{\smallskip}
Line ratios & Region 1 &  Region 2 & Region 3 & Region 4 & Region 5 \\
 & 29\farcs3 & 54\farcs4 & 18\farcs4 & 85\farcs6 & 83\farcs7 \\
\noalign{\smallskip}
 \hline
\noalign{\smallskip}
$[$N II$]$ $\lambda$ 6583/ H$\alpha$            & 0.76 $\pm$ 0.05 & 0.70 $\pm$ 0.08 & 0.60 $\pm$ 0.10 & 0.21 $\pm$ 0.11  & 0.15 $\pm$ 0.06\\
$[$S II$]$ $\lambda$ 6716 + 6731/ H$\alpha$     & 0.99 $\pm$ 0.09 & 0.88 $\pm$ 0.14 & 0.80 $\pm$ 0.18 & 0.15 $\pm$ 0.13  & 0.21 $\pm$ 0.07\\
$[$S II$]$ $\lambda$6716/$[$S II$]$ $\lambda$6731 & 1.46 $\pm$ 0.18 & 1.52 $\pm$ 0.33 & 1.34 $\pm$ 0.44 & $\ga$ 1.3 & $\ga$ 1.5 \\
\noalign{\smallskip} \hline
\end{tabular}
$$
\begin{list}{}{}
\item[$^{\mathrm{a}}$]{Lower limits to the $[$S II$]$
$\lambda$6716/$[$S II$]$ $\lambda$6731 ratio are provided assuming
a 1$\sigma$ upper limit to the flux of the $\lambda$6731 line, if
the line is not detected.}
\end{list}
\end{table*}

\section{Discussion}
In all of the above-mentioned 5 regions the estimated ratio of the
[\ion{S}{ii}] emission lines indicates low electron densities,
N$_{\rm e}$ $<$100 cm$^{-3}$. 

No proper motion, neither of localized nor general nature could be
detected for the CCN, neither by comparing with the POSS (where
the resolution however is poor) nor via a comparison of the NOT
$R$ image (Nov. 1999) with an H$\alpha$ + [\ion{S}{ii}] Tautenburg
image (Oct. 2005). In the case of the latter comparison an
astrometric calibration on the basis of USNO2 was carried out. If
there had been displacements of $\ge$3 NOT pixels (i.e.
$\ge$0\farcs56), we would have noted that. At the CCN's distance
of 150 pc, this corresponds to an upper velocity limit of $\sim$70
km s$^{-1}$.

\subsection{Arc and CCN: projection or connection?}
The [\ion{N}{ii}]$\lambda6583$/H$\alpha$ as well as the
[\ion{S}{ii}]$\lambda6716+6731$/H$\alpha$ emission-line ratios 
gave us a surprise: they
vary from values typical of shock-excited gas in the CCN filaments
(regions 1 to 3) to values typical of photo-ionized gas in the arc
(regions 5 and 4). This assertion can best be examined by using a
diagnostic diagram, like the one presented in Fig. 7 of
\citet{gm91}. There, the CCN falls in the area of SNRs and
HH-objects, while the arc occupies a position within the borders
of \ion{H}{ii} regions. A closer look at the data in Table 1
furthermore reveals that both the [\ion{N}{ii}]/H$\alpha$ ratio
and the [\ion{S}{ii}]/H$\alpha$ ratio have a gradient: the closer
in projection to the arc, the lower the values. Nevertheless there
is a sharp jump as soon as the northern border of the CNN has been
passed.

Although both this gradient and the above-mentioned
H$\alpha$/[\ion{S}{ii}] image support a physical link between the
CCN and the arc, we searched for a further proof for this
connection. Taken alone, the arc could, in principle, represent
the brightest portion of a $\sim20\arcmin$ shell filled with
low-ionized gas, as frequently observed in highly evolved
planetary nebulae or in \ion{H}{ii} regions. Having in mind that
the interstellar extinction is small or negligible in this
direction \citep{zw97}, we examined on POSS red-sensitive and
blue-sensitive copies the whole area of $\sim20\arcmin$ radius for
any blue or bluish star. We found none. An \ion{H}{ii} region can
easily be excluded, since an object of this kind can hardly be
farther away from us than a few kpc (note the Galactic coordinates
of $\ell \approx 197\degr$, $b \approx -38\degr$) and its ionizing
source(s) would, even in the case of a B3V star ($M_{\rm V}$ =
$-1.5\pm0.5$ mag), be of quite high apparent brightness.

Highly evolved planetary nebulae (PNe) of this size, morphology,
and spectral characteristics are, however, known. Examples are
IW\,1, S\,176, and WDHS\,1 \citep{tk96}.  The intrinsically
largest have maximum linear diameters of a few pc. If we assume 3
pc as the maximum, a $\sim20\arcmin$ large PN would be located
about 500 pc distant from us. The central stars of highly evolved
PNe interestingly have closely similar absolute visual magnitudes:
According to \citet{p05} $\langle M_{\rm V} \rangle$ =
7.05$\pm$0.25 mag with an intrinsic scatter of order $\pm$0.5 mag.
In the case of interstellar extinction $A_{\rm V}$ = 0 mag, even
at 500 pc the (strikingly) blue central star of such a PN would
have an apparent magnitude between 15 and 16 mag in $V$. To sum
up, we may hence take it for granted that the CCN and the arc are
in fact two parts of one object.

\subsection{A shock-cloud interaction}
\citet{zw97} suggested that the CCN is a small cloud being
over-run by the slow moving SNR shock wave responsible for the
OEB. We try to substantially add to this conclusion and its
details below. The broad arc, recognized to belong to the CCN,
will thereby be of importance, and we will show that rather the
CCN is a part, better an appendage, of the arc - and not vice
versa.

The above authors found the CCN to be projected onto the centre of
the OEB \ion{H}{i} shell, best seen by plotting the position of
the CCN on the maps of \ion{H}{i} emission in various velocity intervals 
in Fig. 8, particularly Fig. 8d, of \citet{bh95}. The same figure also shows
that one does not deal with an ideal spherical expanding shell,
but with a set of filaments and loops. Hence, even looking at the
projected centre of the shell, one must not expect that the
approaching plane of the shock front is perpendicular to the
line-of-sight. This made us think that the arc and its CCN come
from an originally spherical (or ellipsoidal) cloud, over-run by a
shock front which is oblique with respect to the line-of-sight.

By taking into account results of simulations from the literature
\citep{sn92, km94, op05, bk06, pf02, nmkf06}, we suspect that we deal with
the following situation: The slow (about 40 km s$^{-1}$) shock has
first hit the (perhaps spherical) cloud on the side facing the
OEB, has moved through it and swept over it, producing vortices
and various other instabilities at the rear of the cloud. This
means that the broad arc is nothing else that the still rather
intact original forepart of the interstellar cloud and the CCN is
the unscrewn rear of the cloud. In other words, the system arc +
CCN is inclined to the line-of-sight, where the southern part of
the CCN is closest to us, the northern one farther away, connected
to the remnant of the cloud, which is now a kind of 'cap', closest
to the OEB.

A closer look at the results of published simulations  and a
comparison with our nebulae is very encouraging, since the arc +
CCN can be found to be represented quite well by early to middle
phases in shock-cloud interactions: We particularly refer to Fig.
6 in the experimental work of \citet{kb03} or to Fig. 6 in
\citet{pf02}; the latter figure rather looks like a sketch of the
arc + CCN - with a cap, a filamentary stream of gas heading away
from the cap, a narrowing and then a widening of the filaments. 
In addition, recent simulations considering the realistic case of 
an interaction 
of a planar shock with a smooth boundary interstellar cloud \citep{nmkf06}
result in a density distribution of the shocked cloud at the early
stages of its evolution (their Fig. 2, panels (c) and (d)) 
that closely resembles the observed structure of the system arc + CCN
(although from a different viewing angle). 
At early phases of an interaction, the  surviving (fore)part of the
original cloud did not undergo substantial lateral expansion.
Hence, we can estimate the true linear radius of the cloud,
assuming spherical shape and a lateral expansion of, say, one
third compared to the initial dimension.

The observed radius of 10\arcmin \, corresponds, at D = 150 pc, to
0.44 pc, that is the original radius was $\sim$ 0.3 pc, and hence the
volume of the spherical cloud was 3\,10$^{54}$ cm$^3$. We assume
uniform filling with (originally neutral) gas and further assume
two densities, 90 cm$^{-3}$ and 9 cm$^{-3}$. We note that the
ambient density around the OEB has been found to be rather high,
0.9 cm$^{-3}$ by \citet{bh95}. The higher of the two above
densities accounts for the expected value of the density 
contrast $\chi$ between the cloud
and the ambient intercloud medium, which is of
the order 10$^2$ for cold atomic gas ($T$ $\approx$ 10$^2$ K)
embedded in either the warm neutral medium or the photo-ionized
warm medium ($T$ $\approx$ 10$^4$ K) \citep{kb03}. The lower of
the two values however appears to be perhaps more realistic given
the location away from the plane of the Galaxy and the very small
size of the cloud \citep{ni06}. By assuming \citep[as in][]{zw97} 
a molecular weight $\sim$1.24, we then find a total mass for the 
original spherical
cloud of $\sim$0.3$M_{\odot}$ (or $\sim$0.03$M_{\odot}$) for the
two densities from above. Hence, the shock front of the OEB has
been caught red-handed interacting with an interstellar cloud
of considerably small size and mass.

Why is the arc photo-ionized, but the CCN shock-excited? We
suggest that the arc, which actually seems to be a spherical cap,
is exposed to ionizing photons from the hot gas filling large
parts of the OEB and gas at intermediate temperatures, $T
\sim10^5$ K, that exists between the hot bubble interior and the
surrounding material \citep{ke06}. We also cannot exclude that
there is some UV photon contribution from the Ori OB1 association.
Anyway, the cap appears to act like a protective canopy or screen
and is mildly permeable - note the gradient in Table 1. We believe
that there must also some shock-excitation be present in this cap,
but this is masked by the photo-ionization.

How long can the cloud survive? Practically all theoretical
studies show that small isolated clouds will be destroyed after a
few units of cloud crushing time. This is the timescale for the cloud
to be crushed by the shocks moving into it and is defined
\citep{kb03} as $t_{\rm cc}$ = $\chi^{1/2} a_{\rm 0}$/$v_{\rm b}$,
where in addition to $\chi$, the cloud-intercloud density
contrast, $a_{\rm 0}$ is the original radius of the cloud and
$v_{\rm b}$ is the velocity of the shock wave. Two of these
parameters, $a_{\rm 0}$ and $v_{\rm b}$ are rather well known, and
we take both $\chi$ =  10$^2$ and  $\chi$ = 10 from above. Then,
for $\chi$ =  10$^2$, $t_{\rm cc} =7.1\,10^4$ yr, and for $\chi$ =
10, $t_{\rm cc} = 2.2\,10^4$ yr. In the latter case, our preferred
one, the cloud would hence be completely disrupted after about
10$^5$ years, a period which is short compared to the total age of
the OEB of up to a few 10$^6$ years \citep{bh95}.

An alternative possibility to the scenario described above 
would be that the CCN is just an isolated old SNR in the vicinity
of the OEB. However, we do not favour this alternative point of view
because of the structure of the CCN/canopy system and because of its
radial velocity, which indicates it is a local phenomenon, 
consistent with OEB being the cause for the features.
Furthermore, as already noticed by \citet{zw97}, no radio or
X-ray sources have been detected at its location.
Accurate velocity measurements based on a long-slit high-dispersion
spectrum, not yet available to us, would allow to definitely discard 
(or otherwise accept) this alternative interpretation.
 
To sum up, the CCN and its cap appear to represent an example -
unrivalled in closeness and clarity  - for an early to medium
stage in the destruction of an isolated cloud over-run by an
evolved SNR and thus may deserve further study.

\begin{acknowledgements}
The authors are grateful to F. di Mille for having carried out the
spectroscopic observations at the Asiago Observatory. 
This work
was supported by the Austrian Science Fund (FWF), projects P15316
and P17772.
\end{acknowledgements}


\begin{thebibliography}{}
\bibitem[Boumis et al.(2001)]{bd01} Boumis, P., Dickinson, C., Meaburn, J. et al. 2001,
\mnras, 320, 61
\bibitem[Brown et al.(1995)]{bh95} Brown, A. G. A., Hartmann, D.,
\& Burton, W. B. 1995, A\&A, 300, 903
\bibitem[Byun et al.(2006)]{bk06} Byun, D.-Y., Koo, B.-C.,
Tatematsu, K., \& Sunada, K. 2006, \apj, 637, 282
\bibitem[Garcia-Lario et al.(1991)]{gm91}
Garcia-Lario, P., Manchado, A., Riera, A., Mampaso, A., \&
Pottasch, S. R. 1991, A\&A, 249, 223
\bibitem[Guo et al.(1995)]{gb95} Guo, Z., Burrows, D. N., Sanders,
W. T., Snowden, S. L., \& Penprase, B. E. 1995, \apj, 453, 256
\bibitem[Heiles et al.(1999)]{hh99} Heiles, C., Haffner, L. M., \&
Reynolds, R. J. 1999, ASP Conf. Ser. 168: New Perspectives on the
Interstellar Medium, 211
\bibitem[Klein et al.(1994)]{km94} Klein, R. I., McKee, C. F., \&
Colella, P. 1994, \apj, 420, 213
\bibitem[Klein et al.(2003)]{kb03} Klein, R. I., Budil, K. S.,
Perry, T. S., \& Bach, D. R. 2003, \apj, 583, 245
\bibitem[Koo et al.(2006)]{kk06} Koo, B.-C., Kang, J.-H., \&
Salter, C. J. 2006, \apjl, 643, L49
\bibitem[Kregenow et al.(2006)]{ke06} Kregenow, J., Edelstein, J., Korpela, E. J. et al.
2006, \apjl, 644, L167
\bibitem[Landolt(1992)]{la92} Landolt,  A. U. 1992, \aj, 104, 340
\bibitem[Mac Low \& Klessen(2004)]{lk04} Mac Low, M.-M., \& Klessen,
R. S. 2004, Rev. Mod. Phys., 76, 125
\bibitem[Meaburn(1965)]{mea65} Meaburn, J. 1965, Nature, 207, 179
\bibitem[Meaburn(1967)]{mea67} Meaburn, J. 1967, Zeit f. Astr., 65, 93
\bibitem[Nagashima et al.(2006)]{ni06} Nagashima, M., Inutsuka,
S.-I., \& Koyama, H. 2006, \apjl, 652, 41
\bibitem[Nakamura et al. (2006)]{nmkf06} Nakamura, F., McKee, C. F., 
Klein, R. I., \& Fisher, R. T. 2006, \apjs, 164, 477
\bibitem[Orlando et al.(2005)]{op05} Orlando, S., Peres, G.,
Reale, F., et al. 2005, A\&A, 444, 505
\bibitem[Patnaude \& Fesen(2005)]{pf05} Patnaude, D. J., \& Fesen,
R. A. 2005, \apj, 633, 240
\bibitem[Phillips(2005)]{p05} Phillips, J. P. 2005, \mnras, 357,
619
\bibitem[Poludnenko et al.(2002)]{pf02} Poludnenko, A. Y., Frank,
A., \& Blackman, E. G. 2002, \apj, 576, 832
\bibitem[Seibert et al.(2006)]{ms06} Seibert, M., et al. 2006, in
preparation
\bibitem[Stone \& Norman(1992)]{sn92} Stone, J. M., \& Norman, M.
L. 1992, \apj, 390, L17
\bibitem[Tweedy \& Kwitter(1996)]{tk96} Tweedy, R. W., \& Kwitter,
K. B. 1996, \apjs, 107, 255
\bibitem[van Dokkum(2001)]{vd01} van Dokkum, P. G. 2001, \pasp, 113, 1420
\bibitem[Zanin \& Weinberger(1997)]{zw97} Zanin, C., \&
Weinberger, R. 1997, A\&A, 324, 1165

\end{thebibliography}
\end{document}